%% file: sample-sigconf.tex
\documentclass[manuscript, nonacm]{acmart}
 
\def\BibTeX{{\rm B\kern-.05em{\sc i\kern-.025em b}\kern-.08emT\kern-.1667em\lower.7ex\hbox{E}\kern-.125emX}}
    
\copyrightyear{2021}
\acmYear{2021}
\setcopyright{None}
\acmConference{arXiv Preprint}{November}{2021}

\pdfoutput=1 
\usepackage[linesnumbered, ruled]{algorithm2e}
\usepackage{array}
\usepackage{tabularx}
\newcommand{\rev}[1]{{\color{black} #1}}
\begin{document}

%
\title[The Other Side of Black Screen: Rethinking Interaction in Collaborative Programming]{The Other Side of Black Screen: Rethinking Interaction in Synchronous Remote Learning for Collaborative Programming}

\author{Tahiya Chowdhury}
\affiliation{
\institution{Electrical and Computer Engineering, Rutgers University}
  \city{New Jersey}
  \country{USA}}
%





%
\renewcommand{\shortauthors}{T. Chowdhury et al.}

\input{abstract}


 

%

%

\maketitle

\input{maintext}

\bibliographystyle{ACM-Reference-Format}
\bibliography{sample-base}

\end{document}

%% file: abstract.tex
\begin{abstract}

Collaborative learning environments such as programming labs are crucial for learning experiential hands-on skills such as critical thinking and problem solving, and peer discussion. In a traditional laboratory setting, many of these 
skills can be practiced through natural interaction (verbal, facial) and physical co-location. However, during and after a global pandemic, these learning practices cannot be exercised safely in in-person settings any longer and thus need to be re-imagined for a remote learning environment. As discussions spur about effective remote learning practices, there is an urgency for identifying the unique needs demanded by both students and instructors under different learning environments. \textit{How can we design remote learning to offer broadly accessible learning, by drawing in-person practices and combining them with the power of remote learning solutions?} In this 
case study, we present observations of in-person and online versions of 2 introductory programming courses offered before and during the COVID-19 pandemic. Our observations reveal certain user needs and interaction practices under 5 themes that are unique to students' prior experience with the curriculum and academic level. We find that the current online video-conferencing platforms cannot foster collaborative learning among peers, lacks learning ambiance and spontaneous engagement between students and instructors. Based on our findings, we propose design recommendations and intervention strategies to improve current practices in synchronous remote learning that can facilitate a better learning environment, particularly for introductory lab courses. 

\end{abstract}

%% file: maintext.tex
\section{Introduction}

COVID-19 has been acknowledged as a disrupting factor of everyday human lives around the world. Since the virus has been declared as a pandemic by World Health Organization \cite{cdc_july}, schools and universities around the world shifted to remote learning to avoid the risk involved in resuming in-person classes, which poses several challenges. 
First, delivering learning materials designed for in-person teaching through digital tools is incapable of achieving the same level of engagement and learning outcomes \cite{coppola2002becoming}. Second, travel advisories and border closure forced students around the world to stay in their home countries and rely on remote learning to fulfill academic expectations \cite{nature}.
Lack of firm standardization in online teaching and varied level of access to communication technology creates a digital divide between students and accessible education. \rev{Third, lecture-based courses can be transferred to remote learning through video-conferencing tools while for laboratory courses such as programming and robotics, students can fall behind due to the absence of collaborative learning and social engagement in a laboratory environment \cite{Sun2}.}

As the pandemic hit, many universities were compelled to fast-track on this \textit{in-person to online} transition process and adapt to available 
technology solutions. As an effort to prepare future remote instructors, researchers have surveyed and analyzed the experience of both instructors and students on this transition to identify challenges and design opportunities \cite{Bai, dew2020student, shivam2020remote, fox2020teaching, remote_practical, Roberta, Chola}. 
Like other disciplines, computer science courses also faced significant challenges to ensure quality delivery methods, performance evaluation, and classroom interaction to provide a learning experience that can elicit the positive aspects of an in-person learning environment \cite{Crick}. While there are studies investigating these aspects for remote education during the pandemic \cite{goodman2020learnapplyreinforceshare, Ricks, Crick}, we know little about the obstacles students in introductory lab courses with no prior programming experience face in remote learning.

Let us consider, Alice\footnote[1]{Pseudonyms are used to preserve anonymity.}, a freshman student taking an introductory programming course A. Course A is offered in-person and involves a co-requisite lab with weekly lab assignments. When Alice receives the first lab assignment, they spend three hours trying to solve the problem set and are encountered with a bug in their code. If we assume Alice can navigate across sites such as StackOverflow, we can expect that they can fix their bugs. Alternately, they might plan to reach out to the professor or teaching assistants during office hours or form study groups to study concurrently in university computer labs and libraries which foster collaborative learning. Now, if Alice's university has transferred to online teaching, \textit{what would the workflow of her problem solving and collaborative learning look like? Specifically, what transformation would take place
in the lab experience of students in introductory programming courses?}

\begin{figure*}[ht]
  \centering
  \includegraphics[width=0.90\linewidth]{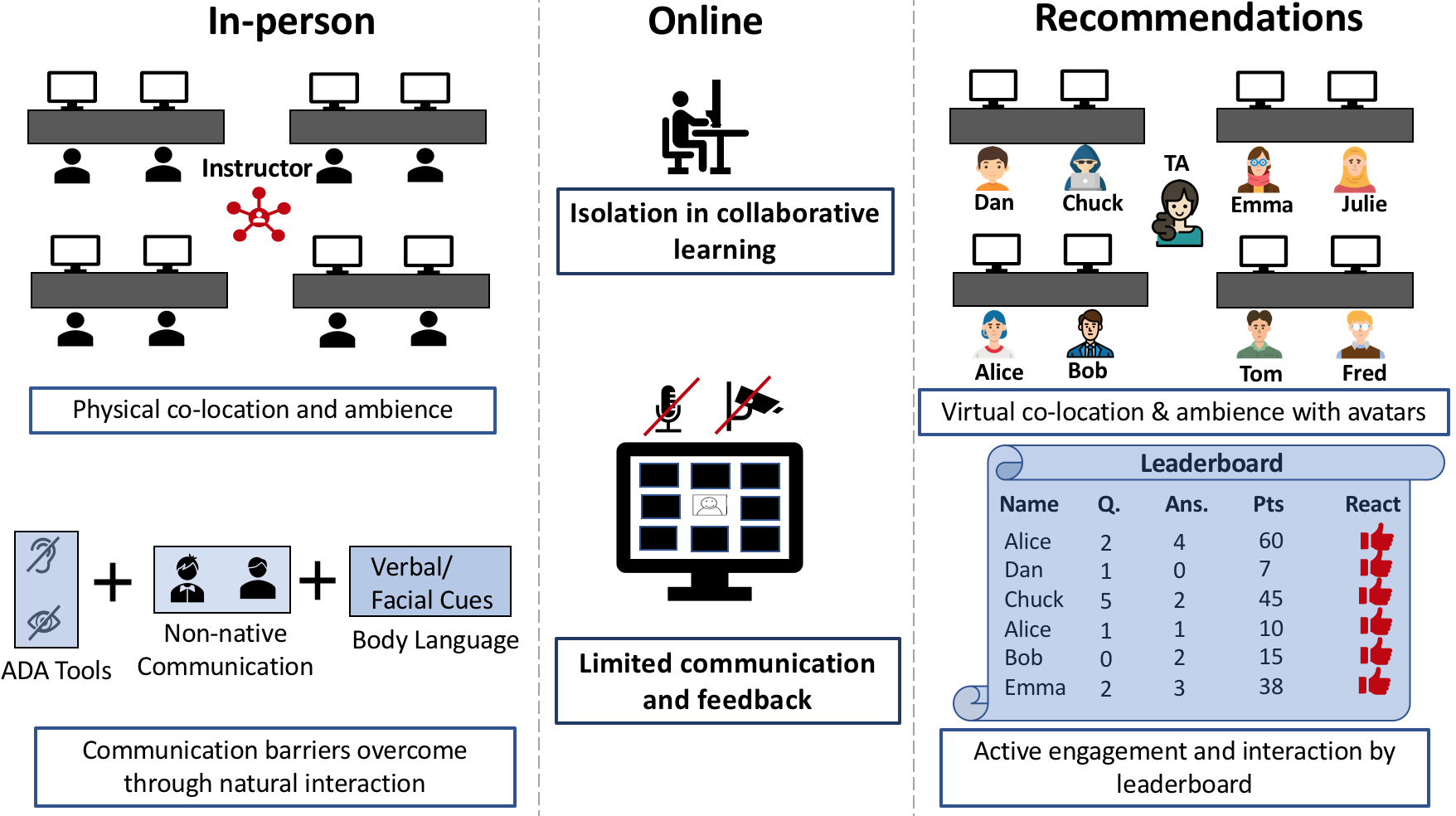}
  \caption{Challenges identified from our case-study of in-person and online lab offerings. We propose intervention recommendations based on our findings (right). Virtual co-location and ambiance created by using avatars have the potential to foster a better social presence among students. Student leaderboard containing questions, answers, and rewarded points to each student can establish better engagement and active participation among them.}
  \label{fig:design-recP}
\end{figure*}

To answer this, we perform a case study of 2 introductory computer programming courses offered at a large, public north-American university before and during the pandemic. In this work, we present first-hand, in-situ observations of the first author, an instructor for both courses to analyze interaction and collaboration practices in introductory programming courses in both versions based on the observation notes documented in 72 hours of lab sessions and 27 hours of office hours. 
The objective of this case study is to observe and analyze student and TA workflow under each version to identify unique needs and interaction patterns to inform design recommendations. Specifically, our study provides insights to the following questions:

\textbf{Q1:}\textit{What is the workflow of problem-solving and collaborative learning, in in-person and remote settings?}

\textbf{Q2:} \textit{What interaction practices are observed in in-person and remote labs and how they are different?} 

\textbf{Q3:} \textit{What design solutions over existing remote learning technology can offer a more complete, interactive experience that resembles existing interaction exercises?}



Our paper makes three key contributions to the computing education research community. First, we provide an empirical understanding of various interaction practices in two synchronous laboratory courses under both in-person and remote settings.
Prior works  \cite{Bai, dew2020student, shivam2020remote, fox2020teaching, remote_practical, Roberta, Chola, Ricks, Crick} focus primarily on asynchronous and hybrid learning with data collected from self-reported surveys and interviews. This leaves open questions of 
empirical understanding on how various interaction practices transition from in-person to remote laboratories and to what extent the transition affects student learning experience. Second, we use our findings from observing lab workflows in the two versions to delineate the factors influencing student learning; in particular, the role of accessibility, culture, ambience, academic level, and prior programming experience.  Our analysis reveals 
five major themes based on the interaction practices observed in these courses:
collaborative learning, engagement, accommodation and inclusion, culture and communication, and attention. Finally, we conclude with practical intervention strategies to support future synchronous remote learning and design recommendations to improve the experience of collaborative education. \rev{Figure~\ref{fig:design-recP} summarizes the observations and recommendations based on our study.}

\section{Background and Related Work}

The work presented in this paper is informed by research on the role of interaction in remote learning, collaboration in distance, as well as remote learning strategies employed for both Massive Open Online Courses (MOOCs) and synchronous courses during the COVID-19 pandemic.

\subsection{Role of Interaction in Remote Learning}

Researchers in education technology and distance education have been exploring interaction strategies and practices in a remote learning classroom for decades. Moore et al. \cite{moore1989three} described three interactive elements for effective remote learning that involve interaction with the content, instructor, and students.  Interaction with the content is a form of communication of the student with written and multimedia format to ensure efficient learning.
Swan et al. \cite{swan2001virtual} showed that students encounter difficulty and
become overwhelmed with self-discipline and the firm framework needed for the successful application of remote learning. Marcus et al. \cite{marcus2003communication} associated the absence of a learning atmosphere and a lack of direct interaction between student and instructor as primary reasons for student dissatisfaction with remote learning. Interaction between student-instructor has a critical connection with learning outcomes and student engagement as found in \cite{swan2001virtual, moore1989three}. The communication gap induced by the physical distance between them requires specific intervention strategies, such as increased dialogue to compensate for verbal and physical cues missing in remote settings \cite{battalio2009success}.

Communication among students to share ideas, to compensate gaps in understanding, and to disagree with others' ideas is an important need in the learning process \cite{picciano2002beyond, haythornthwaite2006facilitating}. In remote learning, this social presence and interaction among peers are offered through discussion forums, video-conferencing tools, and chat rooms. Collaborative learning strategies such as pair programming can help students overcome isolation, encourage brainstorming, and foster an empathetic, inclusive environment as highlighted in \cite{lai1997computer, trentin1998computer, Gray}, but its benefits are shadowed by lack of spontaneity and perceived physical distance in a remote setting. Synchronous remote learning as practiced in a programming laboratory demands for more novel, cheaper, yet effective method to generate engagement and discussion among peers.
Effective interaction strategies improve student learning experiences \cite{schreiber1998distance, wright2000critical}, however, many questions about our understanding of these interactions for synchronous remote learning remain unexplored. 
While synchronous remote teaching and its effect on student-teacher interaction were studied in the literature \cite{coppola2002becoming, bower2015design, stewart2011students}, 
comparative evaluation based on in-class observations for both in-person and remote teaching is scarce. Our work informs a detailed step-by-step understanding of interaction patterns in remote learning, with a focus on introductory programming courses.

\subsection{Remote Collaboration and Interaction}

Collaboration in distance for professional, academic, and social context have attracted the attention of numerous researchers in computer-based collaborative work. Hollan et al. \cite{Hollan92} argued that it is infeasible to afford the richness and variety of interactions when collaborating in the distance by mere imitation of its physically proximate counterpart. This is confirmed by Olson et al. \cite{Olson2000, Olson95} where they outlined that the success of remote collaborative work is tightly coupled with identifying useful tools to re-imagine interaction patterns through virtual co-location. Several works have been done along this line to approximate physical co-presence through spatial transformation \cite{Hauber}, collaborative media curation and sharing \cite{Hamilton}, and audio immersion \cite{Mehrotra} for video-conferencing tools designed for professionals.

In remote learning, there are unique needs that differ by students' academic, socio-economic, and communication levels because, as pointed out by Olson et al., \textit{distance still matters}. Prior works on individuals' struggle to work alone remotely identify social support, visible evaluation and feedback mechanism, shared personal identity, and diverse peer discussion as useful coping strategies to build community in remote learning \cite{Kulkarni, Koehne, Gumienny, Sun1, Sun2}. However, the challenges experienced in student and teaching staff workflow for introductory programming laboratories and its consequences in learning outcomes are not reflected in surveys and interview responses from participants. While there are intervention strategies in the literature that have tried to balance student learning outcomes with human-centered computing \cite{bower2015design, Malan}, the technological and logistic implications of these techniques can be expensive both in terms of money and staff support.

\subsection{Effective Practices for Remote Learning}

Before the COVID-19 outbreak, much of the work that explored remote learning has focused on Massive Open Online Courses (MOOCS). Zheng et al. analyzed teaching challenges for MOOCS from an instructor's perspective and
recommended better support and technology for effective collaborative teaching \cite{Zheng}  and shared time-zone and a virtual classroom for community formation \cite{Zheng2}. 
As MOOCs struggle to retain student engagement and retention, researchers suggested game mechanics \cite{Hew}, reputation system based on points \cite{coetzee2014should} and incremental improvements in technology \cite{Reich} as effective intervention strategies. Reich et al. suggest incremental improvements for technology-driven solutions for future systems of learning-at-scale \cite{Reich}.
Grudin et al. \cite{Grudin} highlighted teaching style, feedback medium other than facial expression in the video, and establishing an interaction convention common for students of all cultures as practical strategies for synchronous learning.

During the COVID-19 pandemic, instructors and education researchers developed various strategies combined with known remote learning practices with impromptu solutions to achieve learning goals and receive feedback. In a sentiment analysis study on college students during COVID-19, Duong et al. \cite{duong2020ivory} showed that 81.3\% of college students showed dislikes for remote learning. Xu et al. \cite{Xu} found a strong decline in performance for males, beginner-level students, Black students, and students that were already under-performing when classes moved online. 
While self-paced learning and the time-saving ability of remote offering are attractive to students \cite{Chola}, increased social interaction with instructors and peers helped alleviate anxiety and improved mental health \cite{Roberta}.
Researchers have further explored the challenges, benefits, and experiences of both synchronous \cite{chen2020largescale}, and asynchronous \cite{Bai} delivery methods.

Many STEM courses are designed with a laboratory co-requisite with a focus on hands-on training of the students on concepts learned in lectures and these courses were no exception to the transition to online setting. 
Bangert et al. \cite{remote_practical} investigated remote learning experiences for hardware practicals and suggested a combination of pre-recorded experiments and synchronous interactive sessions for achieving original learning outcomes. Ricks et al. \cite{Ricks} introduced self-directed experiential learning for computer programming courses, which assumes students possess prior experience of programming. To help instructors design effective remote laboratory learning, instructors explored challenges experienced in transitioning to remote learning through surveys and interviews for introductory physics and engineering labs \cite{fox2020teaching, shivam2020remote} and introductory physics and astronomy courses \cite{dew2020student}. Our work contributes to this body of literature with a focus on analyzing beginner-level students' workflow on task-level details, complemented by TA workflow, and collaborative learning practices among peers in synchronous remote learning for introductory programming courses.

\begin{table}[]
\centering
    \begin{tabular*}{0.56\textwidth}{l||l||l||l||l}
    \hline
    Course & \textbf{Lab Sessions} & \textbf{Office Hrs.} & \textbf{Male} & \textbf{Female} \\ \hline
    \textbf{Course A}  & 26 & 13 & 55 & 11 \\ \hline
    \textbf{Course B}  & 28 & 14 & 16 & 4 \\ \hline
    \end{tabular*}
   
\caption{Statistics of the two courses studied.}
\vspace{-10mm}
\label{tab:dataset}
\end{table}

\section{Method}

\subsection{Study Setting} 

\textbf{Course.} We perform a case study to compare in-person and remote offering of 2 courses at a large, public university. Course A (introductory programming and algorithm course in C and C++) was offered in Spring 2020, in in-person for seven weeks and online for the remaining six weeks after university moved classes to virtual. Course B was offered fully online as a 7-weeks summer course during Summer 2020 with the same curriculum as Course A. Both Courses A and B have been offered for two years before the COVID-19 pandemic. 
Both courses 
involve weekly lectures accompanied by a programming lab as a co-requisite. The courses cover the implementation of different data structures and basic algorithms (searching, sorting) in C/C++. All computer science and computer engineering students are required to take either of the course offerings and the students are traditionally aged with an average age between 18-22 years with little to no prior programming experience.

\textbf{Lab sessions.} The setting for our observation was the programming lab conducted as a co-requisite for the main course (we will refer to the two labs as Course A and B from here). In both cases, the programming labs were designed for students to work on a series of individual, hands-on assignments and projects on various programming concepts. The first author 
has been an instructor of both courses for two years which provides a unique opportunity to observe the courses and changes that occurred in labs by the transition to remote learning.
Note that both Courses A and B were previously offered in-person in the academic year of 2018-2019, where the lab sessions were held in computer labs. In the academic year of 2019-2020, course A (six weeks) and B were offered completely via the remote learning platform. The in-person version of the courses used Sakai as the primary course management system (CMS). The online version used Webex and Zoom as the video-conferencing tool to hold lab and office hour sessions with Sakai as secondary management support. Table~\ref{tab:dataset} shows the statistics of the two courses studied.

\subsection{Data Collection and Analysis}

The courses provide three kinds of learning environments:

\textbf{Real-time TA feedback sessions.} Students work on their weekly assignments following guided instructions during synchronous lab hours and receive real-time feedback on their work.

\textbf{In-class collaborative learning from peers.} During lab sessions, students form small groups to collaboratively interact, discuss, and learn from peers.

\textbf{Office hours.} Open office hours for students to get additional help and discussion with TA.

The first author, who has been an instructor of both courses for two years, worked in two 80-minute long lab sessions and a single 60-minute office hour each week where students meet synchronously to work and get help on their assignments. Our study includes observations from 54 lab sessions and 27 office-hour sessions for the study. 
The first author has been taking field notes in class and during office hours for the in-person offering of Courses A and B with the primary focus to improve future offerings of the course in pre-COVID time. When classes are moved online, the author saw the unique opportunity to study the transition, continued to collect field notes, and have informal discussions with students to improve the remote learning environment. 
The author served as participant-observer during lab sessions where observation notes included observation of students' workflow in completing an assignment, student's interaction with peers and instructor, class participation, and attendance. During office hours, the author (we will refer to the first author as TA from here) took notes of the type of programming issues faced by the student, student's level of programming efficacy, remarks about ongoing practices during lab, and feedback for improvement.
Based on these student responses and field notes, we compare and analyze student learning experience and interaction patterns in the two versions. 
Note that our focus is not identifying benefits or drawbacks through a comparative analysis of a certain platform, as performed in \cite{Pal}. Rather, we focus on identifying key aspects of student learning and interaction experience in the in-person version and recommend design solutions for the online experience of the course to achieve the same learning objective.


Once collected, the field notes have been iteratively analyzed by reading them repeatedly, coding, and categorizing them until certain higher-level themes emerge through an inductive analysis approach \cite{Corbin}. Once we identify the major semantic themes in the observation notes, we analyze the specific interaction scenarios to identify similarities and differences to outline intervention strategies that can close the gap. We only report anonymized experience and findings obtained in aggregated form following the regulations of the Institutional Review Board of our university.

\section{Experiences and Observations}

In this section, we describe our observations of the lab workflow for both in-person and online formats and identify the unique needs and interactions in both versions.

\subsection{Student Workflows}

\textbf{In-person workflow.} The laboratory sessions take place in a university computer lab with assigned workstations available for students. 
Students are free to work on either the workstation computer or their personal laptops. 
Students can receive TA and peer interaction during the lab session on many instances: 1) when they walk into the lab, they become aware of who is presently attending the lab around the laboratory room at a glance, 
2) nearby students are often the first ones to have a discussion and provide feedback to catch a bug, 2) while some students may work on their own mostly, they reach out to help and engage in ongoing discussion with other students, 3) students group together for a persistent bug and show up in office hour where TA and students can work together to solve bugs.

A student, Bob (pseudonym), walks into the university computer lab. Bob opens up the assignment that was posted on the course website. Bob writes a few lines of code, then encounters a bug or an error. Bob tries to find out the source of the bug on their own, then discusses with the nearby person for debugging suggestions. If other people also encounter the same issue, they try to search on the internet for relevant tips or raise their hands to draw the attention of the TA. Alternately, Bob, Alice, and Chuck encounter the bug and Alice eventually finds a work-around for it. Alice walks up to the blackboard or uses pen and paper to show the underlying process that caused the bug. Additionally, if the bug is encountered not during in-class lab sessions where TA is present, students make appointments for office hours to discuss with TA.


\textbf{Online workflow.} Bob logs into the meeting session as there is no physical lab space. Bob opens up the assignment posted on the course website on their personal computer. Bob writes a few lines of code and encounters a bug. They search on the internet or post in the course chat room with error descriptions and screenshots. Other students can join in the discussion, but there are no neighboring students to discuss interactively. Bob starts sharing the screen to show the error to the \textit{entire class}. Alternately, the TA can assign a private room (breakout room) to discuss the error with Bob only. Note that students in the meeting for lab sessions cannot create breakout rooms among themselves for discussion. 

\begin{table*}[]
\begin{tabular*}{\linewidth}{p{0.26\linewidth}|p{0.325\linewidth}|p{0.325\linewidth}}
 \hline
 \textbf{Themes} & \textbf{Interaction Dissimilarities}  & \textbf{Recommendations} \\
 \hline
 Collaborative Learning   & No side-channel communication between peers & Breakout rooms between student subgroups with screen-sharing feature\\
 \hline
 Engagement &  Passive participation and limited feedback from students & Physical ambiance and interactions with sounds and avatars \\
 \hline
 Accommodation and Inclusion & Limited functionality from screen-reader and closed-captioning features &  Virtual co-location and interaction for a democratized learning\\
 \hline
 Culture and Communication   & Lacks facial cues and body language to support non-native speakers  &  Student leader-board to encourage interaction and participation\\
 \hline
 Attention &  Limited connectivity and isolation hinder attention and communication & Smaller lab sections and discretize lab instructions to provide tangible rewards \\
\hline
\end{tabular*}
\caption{\rev{Interaction dissimilarities under five themes were outlined from our observations of online offerings of two laboratory courses studied and the comparison to their in-person counterparts.} We offer intervention recommendations to facilitate interaction between content, students, and instructor based on our findings.}
\vspace{-6mm}
\label{tab:Recommendations}
\end{table*}

\subsection{TA Workflows}




\textbf{In-person workflow.} The TA needs to be present at least 10 minutes before the scheduled starting time of each lab session (according to department regulation). When TA walks into the computer lab, 
if there are students 
already present and working in the lab, they raise their hand to ask for help regarding any current issue they are facing with the assignment. 
Alternatively, in the case of no students experiencing issues (yet), this small interval of time is utilized by the TA to identify possible areas where students may struggle most and need help with. 

At the beginning of lab sessions, students spend some time understanding the assignment and what they are required to do. This often includes revisiting lecture slides to refresh memory (lab assignments are generally designed based on the topics discussed in lectures that week). 
A majority of students make efforts to write a few lines of code and continue to do so
until they are stuck and ask for help. 
When encountered with an issue, a student, Bob, first consults their neighboring students for debugging suggestions. If they discover that other students are encountering similar issues, that helps to keep their frustration level low as they find out such issues to be normal among their peers. Bob then can start discussing the issue with neighboring students, Alice and Chuck, and try to identify the reason for the bug by searching on the web or solving it on pen and paper. If the bug persists, they walk up to the TA to show their thought process and the current version of code for further discussion. 

Sometimes, it is not uncommon that some students to raise a hand asking for help immediately after the lab starts. These students are often visibly disengaged with the rest of the students and often prefer to work by themselves. While some prefer isolation due to their perception of self-efficacy, many (e.g. students from minority groups, with special accommodation needs or language barriers) find it harder to ask for help from their peers during lab sessions at the beginning of the semester. %
Note that this is an introductory class at the sophomore level and students may not have a chance to form a circle of peers. As the semester progresses, these students often become more comfortable sharing the `intermediate' version of their code with their peers due to being physically co-located and sharing lab time over several sessions. Before they reach that stage, the TA often needs to offer help and reassurance to these students that having trouble with syntax and other minor details is normal for beginner programmers so that they can keep their motivation level intact.

\textbf{Online Workflow.} The TA hosts meetings on a web conferencing platform (Webex and Zoom) for the lab session. The invitation link for the lab session is valid for all sessions during the semester and is shared with the students registered for the course through Sakai and email. As the lab session begins on a video-conferencing platform (Webex and Zoom), the TA views the class (students present in the lab session) from the attendee list. and face views of the students who turned on the camera.
Some of the students join the meeting while keeping both their microphone and camera turned off and work on their own without interacting with their peers or TA through chat. On the contrary, a handful of students join with both audio and video turned on from the start and share their work-in-progress whenever possible. They also tend to take full advantage of the meeting platform's different features such as raise hand, screen share, etc. to mimic physical lab experience and get help with their code whenever needed. 
This group also tends to avoid missing labs (based on attendance and submission records). However, the TA observed that the group that works by themselves asks for help individually through private chat or emails and occasionally appears during office hours (also online).
\

\section{Findings}

In this section, we outline our findings from the analysis under five themes that shape interaction practices in synchronous learning for collaborative programming. We later discuss intervention strategies to address the gaps identified in our findings which are summarized in Table~\ref{tab:Recommendations}.

\subsection{Collaborative Learning}

Learning by collaboration with peers is vital for the success of pedagogical practices such as active learning. A programming laboratory inherently adopts an active learning approach by allowing students to apply concepts learned during lectures to solve practical problems
. When introduced with concepts such as manipulating a doubly-linked list and compare a doubly-linked list with a single linked list, beginner students learn more concretely and develop intuitions through discussion with their peers and TAs using charts or diagrams.

\textbf{In-person.} When a student is stuck on a bug while implementing a certain data structure, they attempt to solve it on pen and paper and to understand the concept. Students can also take turns to express their ideas of the problem to each other to understand the concept. They can repeatedly ask questions to both the TA and neighboring students until the understanding becomes clear. Facial, verbal cues, and direct hand gestures help to communicate the idea and create a sense of engagement that aids learning. 

\textbf{Online.} Based on our observation, a student attempts to solve problems on pen and paper when stuck, but sharing it can only be possible by either showing the written face of the paper towards the camera or using a digital drawing tool. This also means that students need to turn their cameras on to share with all participants in the meeting, which can be intimidating for students who are introverts. We find that most students choose not to communicate their problems at all due to this discomfort. Some rely on private chat to send screenshots or e-mail or choose to wait for individual time with TA during office hour, 
which delays students' progress for the assignment and learning outcomes. The in-class engagement from interaction and discussion between peers disappears from the learning life cycle. Most interactions become confined solely between TA and student and lack of peer interaction becomes prevalent.

\subsection{Engagement}

Unlike lectures, the delivery method of synchronous programming laboratory sessions does not demand students to watch a live lecture by looking at the screen for a prolonged period. Students have the freedom to work on the assignment at their own pace. 
However, the physical awareness of being seated in a lab and discussion can prepare the student mentally for learning ahead.
A sense of social awareness towards other students around and the laboratory environment create an ambiance of engagement.

\textbf{In-person.} Student walks into the lab, which often has specific rules (e.g. food or drink in no-lid containers are not allowed, keep your voice low when working during an open lab hour). These practices create a learning environment for students. Even with the physical ambiance offered in the in-person lab, we observe that some students feel disengaged and spend lab time on social media platforms, browsing, or playing games, particularly those who work by themselves. As Thanh showed in \cite{Nguyen}, students who work together in groups generally succeed in spending lab time efficiently working on assigned work.

\textbf{Online.} In the online lab, there is no barrier between the home and the lab environment. While video-conferencing is convenient (as it is only a click away), it is incapable of providing the 
physical ambiance, a sense of awareness of physically being in a lab. We observe that 90\% of the students keep their camera turned off during online lab sessions, which feels isolating, according to several students (from individual discussion). To avoid unnecessary sounds, all participants' microphones are turned off automatically as they join. About 70\% of the students never turned their microphones on (based on our records of student participation). There is no feedback system available for the instructor during a meeting: \textit{Has the student encountered a bug? Is the student comfortable sharing his/her current version of code for feedback? Is the student actively engaged in attending the lab session or is occupied with something else at the moment?} 
The students need to actively draw attention by either turning on the microphone or sharing the screen, which is challenging for most beginner-level students. As a result, only a handful of 
students engage in discussions while the majority remain silent behind the \textit{black screen} and leave as soon as the meeting ends. There are no after-class social interactions among neighboring students to intrinsically motivate students towards learning through a community-building experience. After several sessions ending in this manner, the majority of students feel disengaged from the class and start attending the sessions without active participation or attention to the ongoing conversation.

\subsection{Accommodation and Inclusion}

The existing practice of education delivery is often designed without proper accommodation tools \cite{baker2019educational}. Despite many ongoing efforts \cite{stefik, schanzer2019accessible}, accessibility is still far from being considered as a norm instead of an exception. 

\textbf{In-person.} We observed 3 students with special accommodation needs (1 student with hearing impairment, 2 with a learning disability) in Course A. They all requested in-class instructions to be written in detail either on the blackboard or posted on the course site at the beginning of the semester. The peer learning process has been absent for the students as students with special needs did not interact with others generally and worked mostly by themselves. These students also asked for help from the TA on rare occasions. However, we observed them readily engage in a discussion when requested by a nearby student for debugging advice. Once the assignment has been completed and submitted on the submission portal, we observe that these students tend to leave the lab room immediately, whereas other students tend to stay longer until the lab hour ends and thus gain more opportunity for peer interactions.

\textbf{Online.} Surprisingly, in the online lab version, the students with accommodation needs interacted more often than they did in the in-person lab. They frequently used the chat feature of Zoom to confirm instructions and ask questions, similar to the findings reported in \cite{Lacher}. While several students expressed a feeling of isolation and lack of interaction that affect their learning environment, the students with accommodation needs performed equal or better when the classes were moved online. However, for students with visual and hearing impairments,
following the instructions have been challenging as they remarked that 
Zoom closed-captioning features are not optimally functional. These students leveraged features offered by web conferencing platforms such as sharing a URL or a screenshot to obtain quick feedback as they work through their assignments.

\subsection{Culture and Communication}

In many undergraduate programming courses, a major portion of the students is international. These students learn English as a second language and their spoken English often possesses accents. Research shows that this can limit students' ability and likelihood to interact with class \cite{wang2017international, Carroll} which decreases with their perceived level of accent. International students are likely to interact more with instructors who they think possess similar cultures and accents.


\textbf{In-person.} Many native speakers are very actively engaged in class discussion whereas most students with accents tend to speak less to avoid getting noticed. Participation comes naturally for the native speakers without worrying about the language barrier. 
Students who ask questions become known to their peers by face in a physical classroom and many students welcome this attention. However, over the course of the semester, many students learn to interact with students outside of their small group (often formed of students of similar culture) and overcome this rigidity towards direct interaction. 
Physical co-existence plays a key role in overcoming this barrier of direct interaction over the course of the semester, which is largely mitigated by body language and visual cues. Furthermore, finding peers in students with similar cultures becomes serendipitous in the in-person setting during the first few sessions. 

Improving verbal communication, as an essential interpersonal skill to develop during undergraduate education, help the students to communicate with TA and instructors. As students become more comfortable sharing their concepts with their peers (noted as struggling students and required more help at the beginning of the semester), we observe that the students start putting more effort into solving the assignment and being part of the ongoing discussion when there is one. %
Consequently, this helps the TA to manage time constraints during a lab session. Programming issues experienced by most students are repetitive and by explaining to a group instead of each student individually, the TA can ensure all students receive support and get their questions answered.

\begin{figure}[h]
  \centering
  \includegraphics[width=0.90\linewidth]{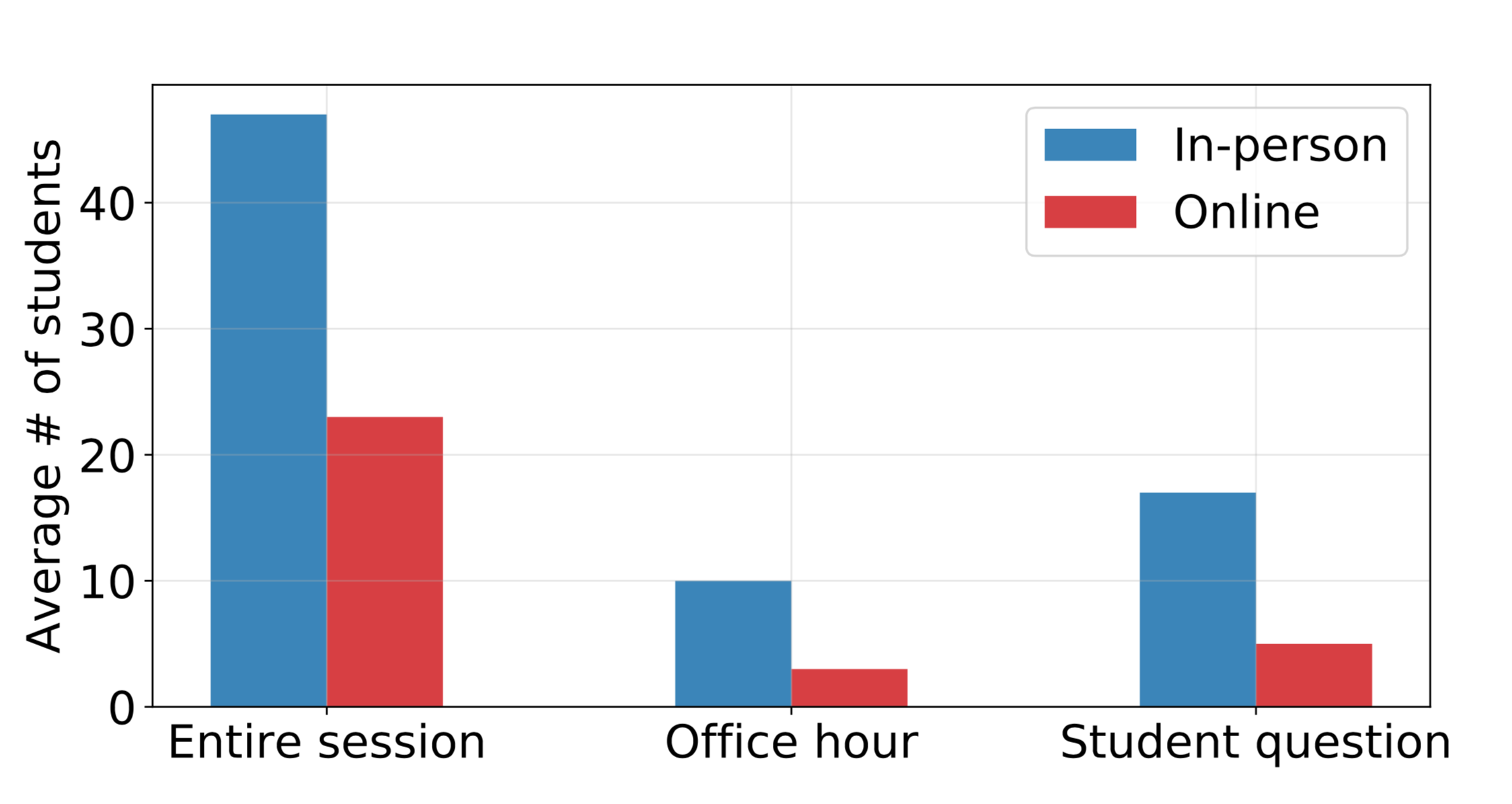}
  \caption{Comparison of student participation (attendance in complete lab session, office hours and asked questions) in in-person and online offering of course A.}
  \label{figure:participation}
  \vspace{-6mm}
\end{figure}

\textbf{Online.} For beginner programming classes, finding peer groups poses severe challenges due to a lack of physical co-existence. Our observation from a fully online and partial online lab shows that students struggled less in a partial online lab than in a fully online one. We argue that this happens as they leverage their already formed peer groups to access peer learning. On the contrary, students with a fully online lab often experience significant challenges without a global view of the class participants. From our study experience, introducing informal discussion forums on platforms such as Slack or Discord can help, but some students remark that they refrain from contributing to programming discussions there due to lack of incentive and due to discomfort in case their suggestions are misunderstood or found to be incorrect later. 

Interaction with the instructor poses another challenge for beginner-level students. Due to a lack of physical co-existence, students find asking questions during sessions particularly challenging. Posting questions through chat allows students to interact without being conscious of their accents. However, concern over the inability to interact without revealing identity may still hinder interaction (each question asked in chat is posted with a name). Some students use a pseudonym and prefer not to set a profile picture. Without a name and a face, the identity of these students becomes a black box for the instructor with no form of meaningful interaction. Furthermore, in-person labs provide opportunities for the TA to prepare mentally when dealing with a student based on the history of interaction between the two. Meeting physically, the TA can retrieve information such as level of effort on previous assignments, accept TA feedbacks, etc., and can help the student more efficiently. However, in the online setting, communication through chat is more spurious and difficult to address for the TA. We observe incidents where multiple students simultaneously post about having issues on the chat such as \textit{'I am getting this error when I do X, what should I do?'}. As the TA starts addressing questions one by one based on the ordering of the posts, students waiting later in the queue can get impatient and start posting more comments about the issues. Due to the nature of the chat-box interface, all comments appear sequentially as they are posted and soon discussion on multiple different issues becomes entangled and difficult to follow.



\subsection{Attention}

A typical student's attention spans around 10 to 15 minutes \cite{wilson2007attention}. Most lectures and labs often span for longer than this duration. As attention deficit is a common issue in a college lecture setting, strategies such as shorter sessions, interactive teaching tools, in-class exercises and quizzes, and a mixture of teamwork and individual learning are used to capture attention. 

\textbf{In-person.} 
When stuck, the attention of a student can easily move on to browsing and playing games instead of working on the assignment. In a physical lab, the TA can walk around the room to see students' progress. 
Alternately, taking roles as driver and navigator (pair programming) stimulates the brain to retain attention and complete the activity without zoning off. As outlined in the previous points, we observe visible improvement in students' motivation and approach to solving the assignment when they discuss it in a group of two or three. Some perceive this as a fun challenge among themselves (we observe remarks such as \textit{Hey Chuck, I think I am going to finish before you today.}). According to some, this is an opportunity to get together, to find a work-around to an error without spending an excessive amount of time on it all by themselves, and avoid losing interest in programming altogether.

\textbf{Online.} There is no feedback about the attention level of the class in an online lab. The effectiveness of active learning strategies cannot be determined from the silence of the turned-off microphones and retaining student attention becomes hard. 
We observed incidents with packet drop where audio and video were gone out-of-sync and noises by unwarranted echos due to poor WiFi as access to digital connective technology and living condition varies among students. 
If such cases of unwarranted noises are not registered early enough, students may lose attention in the lab assignment, start multitasking on the computer, and leave the meeting before it reaches the end. This pattern is observed for both lab sessions and office hours (Figure~\ref{figure:participation}) where attendance for the entire session is reduced by 50\% in both cases. The consequences of this reduction in attendance cause declined student engagement, as the number of students who asked questions is reduced by 65\%  during online sessions compared to the in-person offering.

\section{Design and Intervention Recommendations}

In this section, we present design recommendations based on our observation of interaction and findings that can facilitate a better learning experience for synchronous remote labs. 

\subsection{Re-imagining ambiance and interaction}

Our analysis calls attention to the void of engagement in the remote laboratory experience. The ongoing discussions about remote learning are principally centered around lectures. We perceive that video conferencing tools are incapable of providing physical nearness to fellow peers, often critical to overcoming initial challenges experienced in a beginner-level CS course. The available tools such as Zoom, Webex, and Google Meet are more suitable for one-to-many instruction in a lecture and lacks integrated side-channel communication between subgroups among participants. This is a significant disadvantage to synchronous collaborative learning and can scale beyond programming courses to courses focused on prototyping and iterative improvements \cite{markel2020designing}. This lack of peer interaction in state-of-the-art video-conferencing tools amplifies isolation for students and hampers engagement. The state-of-the-art video-conferencing tools fail to provide ambiance awareness of laboratory learning. 
The pressure to ensure a quiet surrounding and a presentable background provide little motivation for students to interact. 

\textbf{Recommendation.} A solution for side-channel communication among student subgroups can be screen-sharing facilities among a group of attendees. However, having an honor code and provision to disable this feature by the meeting host is needed to avoid its ill-use during exams taken via Zoom. To create an ambiance resembling a physical lab environment, background sound can be introduced. Zoom already provides a virtual background to mask participant background alongside noise reduction. Similar to existing efforts to recreate office ambiance \cite{soundofcolleague} when working remotely, background sounds such as classroom white noise or productivity music can be applied to fill the silence when no dialogues take place.

\subsection{Flowing Attention Inward} 

The lack of richness in communication channels in current remote learning technology fails to imitate the physical counterpart of these experiences. As a result, students feel less inclined to engage in discussions and maintain the role of a silent observer. This is particularly difficult for beginner students experiencing challenging CS concepts for the first time. It is imperative to engage their attention from the preliminary weeks of the course as that critically determines students' ability to maintain progress without lagging behind and the likelihood to succeed in advanced courses.

\textbf{Recommendation.} Use of a classroom leaderboard can be a solution to promote engagement in class. Students will receive virtual rewards for active engagement in discussion in the form of points for asking questions, submitting in-class exercises, sharing the screen, and discussing a solution to the class. The concept of leaderboard draws ideas from the gamers' community and existing social media tools that have proved to be engaging in practice on sites such as Reddit, Quora, etc. However, discussion activities may need to be moderated to ensure their constructive use. 


\subsection{Rethinking Accommodation and Inclusion} We observe in our analysis that although student accommodation needs do not vary significantly between in-person and online versions of the class, 
this is not attributed to better accommodation facilities of the class. Rather, 
accommodation needs are often suppressed from the students' end for the in-person case and thus cannot be addressed by course authority in such case. This is also observed for under-represented minority students, who avoid interacting in class in front of their peers to avoid criticism, intimidation, and discomfort.

\textbf{Recommendation.} Analysis of the online version of the courses reveals that the students with special accommodation needs perform the same or better after class moved online if their explicit request for additional support is fulfilled (to receive all in-class instructions shared with them in written format). Note that due to lack of physical co-location, addressing these needs is only possible when students communicate about them. Lack of physical co-location makes understanding these needs through body language or facial expression impossible.
However, the online class offers a unique opportunity for students with underrepresented minority backgrounds and accommodation needs to interact without worrying about experiencing bias by revealing their identity. To truly democratize the class, we propose that using avatars can create a level-playing field for students with all kinds of backgrounds and accommodation needs. In this way, students can also interact during a Zoom meeting without having to share their faces as this can be a social barrier due to the different conditions among students. However, having a mixture of attendees in their true persona and avatars can impose a new imbalance in interaction.

\section{Limitations and Future Work} 
An important limitation of our study is we primarily rely upon workflow observation during lab sessions and office hours. The first author observed students only in lab sections under their supervision. Although the lab courses were offered with lectures concurrently, they were excluded from our observation and analysis. To address this, future work should focus on data collection from observations made both in lectures and lab courses to understand the overall student learning experience in a remote setting. Another limitation is that the findings are based on one instructor who has been a part of the teaching staff for multiple offerings of both courses. Follow-up research work is needed to perform a large-scale analysis that involves workflow observation aided with data collected by surveys and interviews of students and instructors across multiple courses. Another important aspect to consider is our work did not collect data on student performance to map the effect of remote learning on student performance \rev{(the student population of course B in the two versions are different and hence their performance cannot be compared)}.
Future work can explore student performance in concurrently running and subsequently offered online courses to understand student learning outcomes across courses. Finally, our study is a stepping stone towards understanding the interaction and collaboration practices in remote learning for ability-diverse student groups. Future work can investigate remote learning for students with special accommodation needs to develop a more detailed understanding of their unique needs. 


\section{Conclusion}

In this paper, we present design solutions that can bridge the gap between virtual collaborative lab environments and programming learning outcomes. We outlined several in-person interaction dimensions between content, instructor, and students which cannot be practiced in present remote learning via video-conferencing tools. Based on the following five themes: collaborative learning, engagement, accommodation and inclusion, culture and communication, and attention, we have proposed intervention strategies that can facilitate a better learning environment for both students and teachers. We further propose design recommendations for extending current tools to emanate a more interactive, engaging learning experience. Our analysis raises new opportunities for further empirical research and design of collaborative remote learning that can provide better support for interaction beyond just \textit{being there}.